\documentclass[aps,prl,twocolumn,showpacs,floatfix]{revtex4}
\usepackage{graphicx}
\usepackage{hyperref}
\usepackage{amsmath}
\usepackage{color}

\newcommand{\ket}[1]{|{#1}\rangle} 

\newcommand{\ud}[1]{{#1^{\dagger}}}

\newcommand{\mean}[1]{\langle #1 \rangle}

\begin{document}
\flushbottom
\title{Violation of classical inequalities by photon frequency-filtering}

\author{C. S\'anchez Mu\~noz}
\author{E. del Valle}
\author{C. Tejedor}
\author{F.P. Laussy}
\email{fabrice.laussy@gmail.com}
\affiliation{Departamento de F\'isica Te\'orica de la Materia Condensada
  and Condensed Matter Physics Center (IFIMAC), Universidad Aut\'onoma
  de Madrid, E-28049, Spain.}

\begin{abstract}
  The violation of the Cauchy--Schwarz and Bell inequalities ranks
  among the major evidences of the genuinely quantum nature of an
  emitter. We show that by dispensing from the usual approximation of
  mode correlations and studying directly correlations between the
  physical reality---the photons---these violations can be
  optimized. This is achieved by extending the concept of photon
  correlations to all frequencies in all the possible windows of
  detections, with no prejudice to the supposed origin of the photons.
  We identify the regions of quantum emission as rooted in collective
  de-excitation involving virtual states instead of, as previously
  assumed, cascaded transitions between real states.
\end{abstract}
 \date{\today} \maketitle
\section{Introduction}

Classical descriptions of the electromagnetic field~\cite{loudon80a}
and local hidden variable theories~\cite{bell66a} yield a series of
inequalities that impose an upper limit to the correlations between
two modes and whose violation prove unequivocally the non-classical
character of quantum mechanics~\cite{reid86a}. Among such equalities,
the Cauchy--Schwarz inequality and Bell's inequalities are prominent
examples that have been put to scrutiny in a large and varied set of
platforms.  The Cauchy--Schwarz Inequality (CSI)~\cite{hardy_book52a}
is one of the most important relations in all of mathematics. It
states that fluctuations of products of random variables are bounded
by the product of autocorrelations: $|\langle
XY\rangle|\le\sqrt{\langle X^2\rangle\langle Y^2\rangle}$. When $X$
and $Y$ are quantum observables, however, this relation can be
violated. That is to say, quantum correlations between two objects can
be so strong as to overcome their individual fluctuations in a way
that is unaccountable by classical physics. Bell's inequalities (BI),
on the other hand, refer to the wider problem of the nonlocal
character of quantum mechanics~\cite{aspect13a}. Their violation
decides in favour of quantum theory over local hidden variable
theories. The underlying correlations are well known to power quantum
information processing~\cite{ekert91a}.

The first experimental demonstrations of violation of these
inequalities were realized in the 70s in the radiation of an atomic
two-photon cascade for the CSI~\cite{clauser74b} and in the early 80s
for the BI~\cite{aspect81a,aspect82a}. There has been a
large body of literature confirming and documenting such violations
ever
since~\cite{rowe01a,kuzmich03a,balic05a,thompson06a,sakai06a,marino08a,kheruntsyan12a}.
Most experimental realizations in both cases involve the correlation
of photons of different frequencies emitted in a multi-photon process,
such as atomic cascades~\cite{aspect81a} or four-wave
mixing~\cite{thompson06a,srivathsan13a}.  While in the underlying
theoretical models these photons are attributed to quantum modes
corresponding to specific optical transitions~\cite{reid86a}, the only
physical reality perceived by the measuring devices are the photons
themselves. One can therefore inquire what are the correlations
between photons with a given property---typically, frequency and polarization for CSI and BI respectively---with no theoretical prejudice as to their
origin. In this text, we address this question in a general context
for frequency correlations, but to fix ideas, we will illustrate our
claims on one particular source of photons. To emphasize that the
frequency-correlated photons do not need to be attached to different
modes, we will consider a single-mode emitter. The simplest
non-trivial candidate---resonance fluorescence---is also of great
intrinsic interest and has been a favourite testbed of quantum
optics~\cite{kimble76a}.

\section{Frequency correlations in resonance fluorescence}

\begin{figure}[t]
\begin{center}
\includegraphics[width=\linewidth]{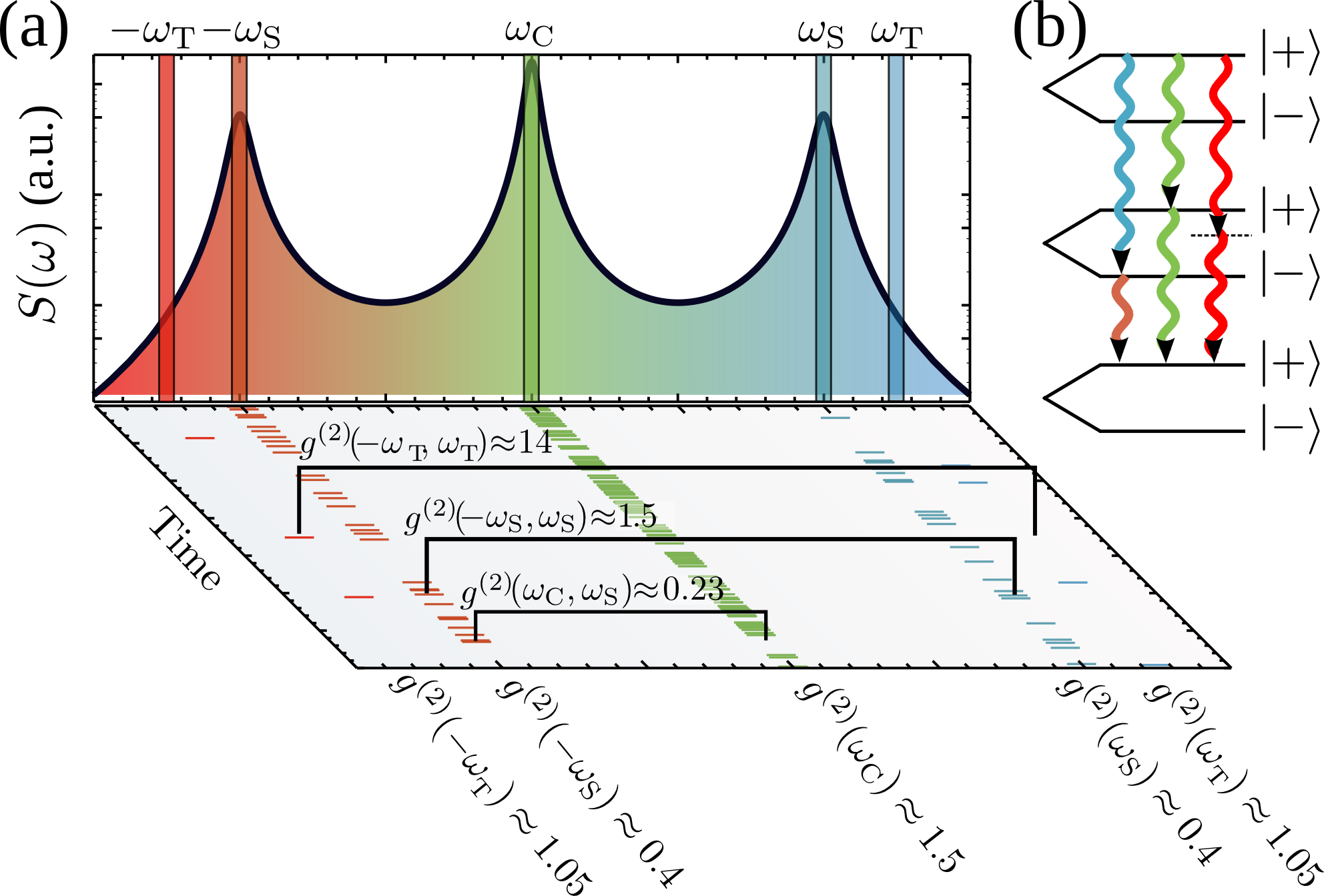}
\end{center}
\caption{(Color online) Violation of CSI and BI by frequency-resolved
  correlations. (a) Spectrum of resonance fluorescence, where
  filtering is illustrated in the tails (T), sidebands (S) and central
  peak (C) of the Mollow triplet. (b) Two-photon de-excitation between
  rungs of the Mollow ladder involve an intermediate real state (blue,
  orange and green arrows) or a virtual state (red arrows). The latter
  type conveys CSI and BI violation. It is found in the flanks or
  between the peaks, where the signal is however weaker.  Parameters:
  $\Omega = 10 \gamma_\sigma$, $\Gamma = \gamma_\sigma$,
  $\omega_\text{S} \approx 2\Omega$, $\omega_\text{T} = 2.5\Omega$.}
\label{fig:1}
\end{figure}

Resonance fluorescence refers to the light emitted under strong
coherent driving by a two-level system
(2LS)~\cite{flagg09a,astafiev10a,lodahl13a}. At high pumping
intensity, the luminescence spectrum splits into three peaks, known as
the Mollow triplet~\cite{mollow69a} (cf.~Fig.~\ref{fig:1}(a)). While
the emission comes from a single mode, $\sigma$, the distinctive
spectral shape calls naturally to question what are the correlations
of---and between---the three peaks. It has been suggested
theoretically~\cite{cohentannoudji79a,apanasevich79a,schrama92a,nienhuis93a}
and established experimentally~\cite{aspect80a,schrama92a, ulhaq12a}
that the photons from the peaks are strongly correlated.
The Hamiltonian for this system reads:
\begin{equation}
 H_0 = \omega_\sigma\sigma^\dagger \sigma + \Omega(e^{-i \omega_\mathrm{L}t}\sigma^\dagger + e^{i \omega_\mathrm{L} t}\sigma)
  \label{eq:hamiltonian}
\end{equation}
with $\omega_\sigma$ the energy of the 2LS and $\Omega$ the intensity
of the field driving it with frequency~$\omega_\mathrm{L}$. With
little loss of generality we will consider resonant excitation:
$\omega_\mathrm{L}=\omega_\sigma$. Dissipation for the emitter is
included in the density matrix formalism as a Lindblad term
$\mathcal{L}_\sigma \rho$ with decay rate~$\gamma_\sigma$ in the
master equation~\cite{breuer_book02a}:
\begin{equation}
\dot{\rho} = -i\left[H_0,\rho\right]+\frac{\gamma_\sigma}{2}\mathcal{L}_\sigma\rho\
\end{equation}
where
$\mathcal{L}_\sigma\rho=2\sigma\rho\ud{\sigma}-\ud{\sigma}\sigma\rho-\rho\ud{\sigma}\sigma$. One
can solve this equation to obtain an analytical expression of the
spectrum featuring the Mollow triplet~\cite{mollow69a,delvalle10d},
which at resonance presents a central peak and two sidebands at
$\omega = \omega_\mathrm{L} \pm \omega_\mathrm{S}$, where
\begin{equation}
  \omega_\mathrm{S}= \Re\{\sqrt{(2\Omega)^2-(\gamma_\sigma/4)^2}\}\,.
\end{equation}

At this point, an ad hoc multiple-mode description is usually enforced
out of the genuine single mode~$\sigma$ which, dressed by the laser,
yields three types of transitions between the dressed
states~$\ket{\pm}$ (cf.~Fig.~\ref{fig:1}(b)). This allows to introduce
three auxiliary modes, associated to the three peaks:
$\sigma_1=c^2|-\rangle\langle+|$, $\sigma_2=cs\left[|+\rangle\langle
  +|-|-\rangle\langle-|\right]$ and $\sigma_3=-s^2|+\rangle\langle-|$
with $s$ and~$c$ two amplitudes~\cite{apanasevich79a,nienhuis93a}.
One can easily compute correlations
$\langle\ud{\sigma_i}\sigma^\dagger_j\sigma_j\sigma_i\rangle$ for
$1\le i,j\le3$ between these modes, that are associated in the
input/output formalism to those~$\langle\ud{a_i}\ud{a_j}a_ja_i\rangle$
of the detected photons with a given
frequency~\cite{gardiner_book00a}.

There are various shortcomings to this approach, which is an
approximation rooted in the physical picture of the dressed atom.
First, the identification of each photon to a given transition based
on its frequency is a simplification. Although infrequent, it happens
that a photon detected at the frequency of a given peak actually
originates from the transition that chiefly accounts for another
peak. When considering regions of overlap, such a misattribution can
become a source of large errors. Second, this approach neglects
interferences between photons that truly are emitted by the same
mode~$\sigma$. Third, modes defined in this way are usually
non-commuting, and therefore correlations at zero delay can yield
different results depending on the order of the
operators~\cite{schrama92a}. Finally, this approach also restricts the
calculation to the three modes thus defined, while one can correlate
any two frequency windows, of various widths and centered at arbitrary
frequencies, not compulsorily at the peak maxima.

\section{Theory of frequency correlations}

To dispense from these approximations and constrains, an exact theory
of frequency-resolved photon detection is required to correlate any
two photons based only on their measured properties, with no
assumption as to their origin or time of emission. The formal
expression for the second-order correlation function between photons
of two different frequencies without recoursing to contrived modes has
been formalized in the late 80s~\cite{knoll86a,cresser87a}. We will
denote it $g^{(2)}_{\Gamma}(\omega_1,\omega_2)$. It provides the
statistics of photons with frequencies $\omega_1$ and~$\omega_2$
spectrally filtered in a Lorentzian window of width~$\Gamma$.  The
resulting integral form turns out to be so awkward, however, that even
in the possession of the expression, there was the need to come back
to the multi-mode approximation to compute it.  In this text, we
recourse to del Valle \emph{et al.}'s theory of frequency-resolved
photon correlations~\cite{delvalle12a} to compute exactly this
measurable property, with no intermediate artificial modes and,
therefore, taking into account all the possible interferences and
indistinguishability imposed by quantum mechanics. This theory
establishes that frequency-resolved correlations of the light emitted
by any open quantum-system are the same as the correlations between
``sensors'' at these frequencies. These sensors are bosonic, commuting
modes with annihilation operator~$a_i$, $i=1,2$, free energy
$\omega_i$ and decay rate~$\Gamma$---accounting for the frequency
linewidth of the sensors---that are weakly coupled to the emitting
mode with a small coupling constant $\varepsilon$. They are included
in the dynamics by the Hamiltonian term $H_S = \sum_i \omega_i
a^\dagger_i a_i + \varepsilon(a^\dagger_i \sigma + a_i
\sigma^\dagger)$ and Lindblad terms
$\frac{\Gamma}{2}\sum_i\mathcal{L}_{a_i}
\rho$~\cite{delvalle12a}. Frequency-resolved correlations are then
computed as:
\begin{equation}
  g_{\Gamma}^{(2)}(\omega_1,\omega_2)= \lim_{\varepsilon\rightarrow 0}\frac{\langle a_1^\dagger a_2^\dagger a_1 a_2\rangle}{\langle a^\dagger_1 a_1 \rangle\langle a^\dagger_2 a_2 \rangle}\,.
\label{eq:g2w1w2}
\end{equation}

With such a theoretical apparatus, a full mapping of the photon
correlations can be obtained. For the case of the Mollow triplet that
we have chosen for illustration, the problem takes the vivid form
pictured in Fig.~\ref{fig:1}. The spectral shape---the triplet---is
represented in log scale with a choice of five frequency windows,
centered at $\pm\omega_\mathrm{T}$ (tails), $\pm\omega_\mathrm{S}$
(sidebands) and~$\omega_\mathrm{C}$ (central peak). A quantum Monte
Carlo trajectory was calculated to simulate the photon-detection
events~\cite{plenio98a} for photo-detectors measuring in these
windows. The emitted photons in a small fraction of the trajectory are
represented with ticks on the projected plane of Fig.~\ref{fig:1}(a).
The intensities vary in each frequency window: there is of course more
signal in the central peak than in the sidebands and more so than in
the tails. What is of interest in quantum optics is the statistical
distribution of, and the correlation between, these photons. The
auto-correlation in a given window, shown in the lower part of
Fig.~\ref{fig:1}(a), gives the statistics of emission of the stream of
photons now defined by their mean frequency and spread. While the
light emitted by the two-level system overall is perfectly
antibunched, one sees that by spectral filtering, one can ``distill''
light with different statistical properties~\cite{delvalle13a},
namely, i) uncorrelated in the tails, ii) antibunched in the satellite
peaks and iii) bunched in the central peak. One can similarly
calculate the cross-correlations between photons from two different
windows, showing this time that photons from the satellites are
positively correlated,
$g^{(2)}_\Gamma(-\omega_\mathrm{S},\omega_\mathrm{S})\approx1.5$,
while photons from one satellite and the central peak are
anti-correlated, with
$g^{(2)}_\Gamma(\omega_\mathrm{C},\omega_\mathrm{S})\approx0.23$. It
is worth noting here that the stronger correlations come from the tail
events, with
$g^{(2)}_\Gamma(-\omega_\mathrm{T},\omega_\mathrm{T})\approx14$ for
the window chosen, and increasing with greater still separations.  The
price to pay for these strong correlations is a correspondingly
vanishing signal. Events are more rare but the strength of their
correlations is increased. This is a general trend.

\begin{figure}[t]
\begin{center}
\includegraphics[width=\linewidth]{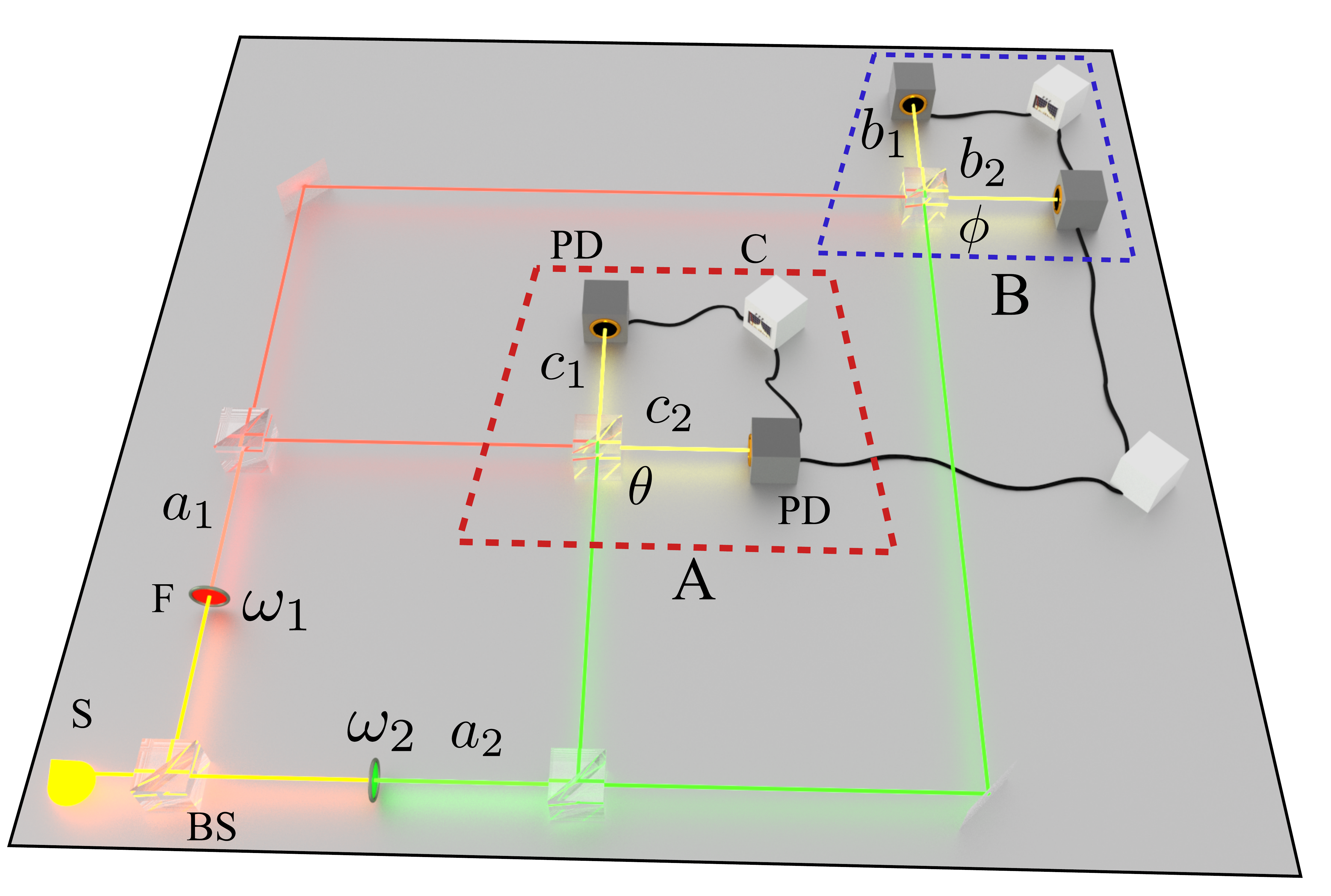}
\end{center}
\caption{(Color online) Test for the violation of Bell inequalities by
  frequency filtering.  A source (S) emits photons in a broadband of
  frequencies. Frequency filters (F) select light at frequencies
  $\omega_1$ and $\omega_2$, described by the operators $a_1$ and
  $a_2$. Recombination at beam splitters (BS) with transmittivities
  given by $\sin\theta$ and $\sin\phi$ gives a total of four output
  beams, which are collected at the photodetectors (PD) and correlated
  with coincidence counters (C). Alice (A) and Bob (B) test
  nonlocality by independently measuring probability of detection at
  the output ports of the two beam splitters, $P^{A_\theta}_\pm$
  and~$P^{B_\phi}_{\pm}$.}
\label{fig:scheme}
\end{figure}

\section{Violation of Cauchy-Schwarz and Bell's inequalities by frequency filtering}

In a quantum optical context, Cauchy-Schwarz and Bell's inequalities
can be expressed through the correlators $\langle a^\dagger_i
a^\dagger_j a_j a_i \rangle$, with $i,j\in\{1,2\}$, of two
electromagnetic modes $a_1$ and $a_2$.  In terms of Glauber's
second-order correlation functions at zero delay
${g^{(2)}_{ij}=\langle a_i^\dagger a_j^\dagger a_j a_i
  \rangle/(\langle a_j^\dagger a_j \rangle \langle a_i^\dagger a_i
  \rangle)}$~\cite{glauber63b}, the CSI reads
$\left[g^{(2)}_{12}\right]^2\leq g_{11}^{(2)} g_{22}^{(2)}$. This can
be expressed in terms of a ratio $R$ that quantifies the degree of CSI
violation:
\begin{equation}
  R = {\left[ g^{(2)}_{12}\right]^2}\Big/\left[g^{(2)}_{11} \,g^{(2)}_{22}\right]
  \label{eq:CSIg2}
\end{equation}
so the CSI takes the form:
\begin{equation}
R \leq 1\,.
\label{eq:CSI-R}
\end{equation}
One can use the definition~\eqref{eq:g2w1w2} for the cross
correlations of Eq.~\eqref{eq:CSIg2} to obtain a degree of CSI
violation for frequency filtered light, $R_\Gamma(\omega_1,\omega_2)$.

The case of BI is less straightforward but can be cast in the same
form. In the CHSH framework~\cite{clauser69a}, one considers
correlated pairs of particles. One of these particles enters an
apparatus where an observable $A_\theta$ is measured while the other
particle enters another apparatus where an observable $B_\phi$ is
measured. $\theta$ and $\phi$ are adjustable parameters of the
apparatuses, i.e., a polarization angle.  The results of each
measurement must be dichotomic, i.e., in each apparatus, the particle
must select one of two possible channels of the observables, providing
values $\pm 1$ (in some units) with probabilities
$P^{(A_\theta)}_{\pm}$ and $P^{(B_\phi)}_{\pm}$,
respectively. Therefore, the measurement in each apparatus yields the
mean values ${\langle A_\theta\rangle =
  P^{(A_\theta)}_+-P^{(A_\theta)}_-}$ and ${\langle B_\phi \rangle =
  P^{(B_\phi)}_+-P^{(B_\phi)}_-}$. As a consequence of this dichotomic
character, the correlation $E(\theta,\phi)=\langle A_\theta B_\phi
\rangle$ between both observables reads:
\begin{equation}
E(\theta,\phi)= P^{(A_\theta,B_\phi)}_{+ +}+P^{(A_\theta,B_\phi)}_{- -}-P^{(A_\theta,B_\phi)}_{+-}-P^{(A_\theta,B_\phi)}_{-+}\,,
\label{eq:E}
\end{equation}
where $P^{(A_\theta,B_\phi)}_{\pm \pm}$ is the joint probability of
measuring $A_\phi=\pm 1$ and $B_\phi = \pm 1$. From this expression in
a local hidden-variable theory, one can derive a BI in the CHSH
form~\cite{clauser69a, reid86a}:
\begin{equation}
  B\leq 2
\label{eq:bell}
\end{equation}
where 
\begin{equation}
B = \left|E(\theta,\phi)-E(\theta,\phi')+E(\theta',\phi')+E(\theta',\phi)\right|.
\label{eq:b-parameter}
\end{equation}

To clarify these concepts, we first consider the case usually discussed in which the particles being correlated are photons and the measurements are done in the polarization degree of freedom. By the nature of the observable, $P^{(A_\theta)}_\pm$ corresponds to the fraction of the total intensity at both output arms of a polarizing beam splitter:
\begin{equation}
  {P^{A_\theta}_\pm=\mean{I^{(A_\theta)}_\pm}/\mean{I^{(A_\theta)}_+ +I^{(A_\theta)}_-}}\,,
  \label{eq:Pa}
\end{equation}
where the adjustable parameter $\theta$ corresponds to the
polarization angle. Correspondingly, the joint probability reads:
\begin{equation}
  P^{A_\theta,B_\phi}_{\pm \mp} = \frac{\mean{I^{(A_\theta)}_\pm I^{(B_\phi)}_\mp}}{\mean{(I^{(A_\theta)}_+ +I^{(A_\theta)}_-)(I^{(B_\phi)}_+ + I^{(B_\phi)}_-)}}\,,
  \label{eq:PP}
\end{equation}
and therefore, we can write $E(\theta,\phi)$ as:
\begin{equation}
  E(\theta,\phi)= \frac{\langle(I^{(A_\theta)}_+-I^{(A_\theta)}_-)(I^{(B_\phi)}_+-I^{(B_\phi)}_-) \rangle}{\langle(I^{(A_\theta)}_+ + I^{(A_\theta)}_-)(I^{(B_\phi)}_+ + I^{(B_\phi)}_-) \rangle}\,.
\label{eq:E-classical}
\end{equation}

This is the typical situation when measuring BI violations for states
of the type:
\begin{equation}
  \label{eq:miejul2202706CEST2014}
  \ket{\psi}=\frac{1}{\sqrt{2}}(a_{1+}^\dagger a_{2+}^\dagger + a_{1-}^\dagger a_{2-}^\dagger)\ket{0}\,,
\end{equation}
where $a^\dagger_{i,\pm}$ is the creation operator for a photon with
polarization $\pm$ along path $i$, i.e., for states that are
entangled. 

In our work, we focus on the correlations from the output of a
dynamical process, that is, we do not restrict to deterministic pure
states~\cite{reid86a} but consider a steady state as an input.  This
means that the intensities $I^{(A_\theta/B_\phi)}_{\pm}$ are no
restricted to unity but can take any positive value. Moreover, we focus on a different scenario that does not involve the polarization degree of freedom, but only two modes states of the type~$\ket{\psi}=a_1^\dagger a_2^\dagger\ket{0}$. If disposing of an emitter that provides such a two-mode output, it is immediate to bring it into an entangled form 
\begin{equation}
\ket{\psi}=\frac{1}{2}(a^\dagger_{1 +}a^\dagger_{2+} - a^\dagger_{1-} a^\dagger_{2-}+i a^\dagger_{1-}a^\dagger_{2+}+i a^\dagger_{1+} a^\dagger_{2-})\ket{0}
\end{equation}
by placing two beam splitters across paths 1 and 2. Subscripts $\pm$ then refer to path instead of polarization. By recombining the four resulting beams in two additional beam splitters with variable transmitivities, that act as the apparatuses measuring $A_\theta$ and $B_\phi$, these states can also violate the BI by following the same line of reasoning as exposed
above~\cite{clauser69a,reid86a}.  $\theta$ and $\phi$ represent in this case the
tunable transmitivities of the two final beam splitters. 

The setup implementing such a scheme
of BI based on frequency filtering is sketched in
Fig.~\ref{fig:scheme}, where the path degree of freedom $1,2$ is associated to the energy degree of freedom $\omega_1,\omega_2$ by using frequency filters. The two possible
channels of detection in each final beam splitter are then equivalent to the
two output ports of the polarizing filters of the conventional case,
and the arguments that led to Eqs.~\eqref{eq:Pa} and~\eqref{eq:PP}
apply similarly. 
In a quantum-mechanical treatment, the modes at the output arms of
the beam splitters are given by:
\begin{align}
c_1 &= \cos\theta a_1 + \sin{\theta}a_2\,, \quad c_2 = -\sin{\theta} a_1 + \cos{\theta} a_2\,, \notag\\
b_1 &= \cos\phi a_1 - \sin{\phi}a_2\,, \quad b_2 = \sin{\phi} a_1 + \cos{\phi} a_2\,,
\label{eq:beam-splitter}
\end{align}
and $E(\theta,\phi)$ takes the form:
\begin{equation}
E(\theta,\phi)= \frac{\langle:(c_1^\dagger c_1-c_2^\dagger c_2)(b_1^\dagger b_1-b_2^\dagger b_2) :\rangle}{\langle :(c_1^\dagger c_1+c_2^\dagger c_2)(b_1^\dagger b_1+b_2^\dagger b_2) : \rangle}\,.
\label{eq:E-quantum}
\end{equation}
We adopt the standard choice of angles that provides the greatest
violation of the inequality: $\theta = 0$, $\phi = \pi/8$,
$\theta'=\pi/4$, $\phi'=3\pi/8$. This yields the following expression
for $B$:
\begin{equation}
  \resizebox{1.\hsize}{!}{$B = \sqrt{2}\left|{\frac{\langle a^{\dagger 2}_1 a_1^2 \rangle + \langle a^{\dagger 2}_2 a_2^2 \rangle - 4\langle a^{\dagger}_1 a^\dagger_2 a_1 a_2\rangle -\langle a_1^{\dagger 2}a_2^2 \rangle -\langle a_2^{\dagger 2}a_1^2 \rangle}{\langle a^{\dagger 2}_1 a_1^2 \rangle + \langle a^{\dagger 2}_2 a_2^2 \rangle + 2\langle a^{\dagger}_1 a^\dagger_2 a_1 a_2 \rangle}}\right|$}.
\label{eq:ansari}
\end{equation}

It is equally easy to formulate these concepts in terms of frequency
correlations than for the CSI. The operators $a_1$ and $a_2$ in
Eq.~\eqref{eq:beam-splitter} can be replaced by the sensor operators
previously introduced and employed into Eq.~\eqref{eq:g2w1w2}, thus
describing the light emitted at the two frequencies $\omega_1$ and
$\omega_2$, as shown in Fig.~\ref{fig:scheme}. Direct application of
Eq.~\eqref{eq:ansari} with these sensors $a_i$, whose finite linewidth
$\Gamma$ is described by their decay rate, provides  $B_\Gamma(\omega_1,\omega_2)$.

\begin{figure*}[t!]
  \includegraphics[width=0.95\textwidth]{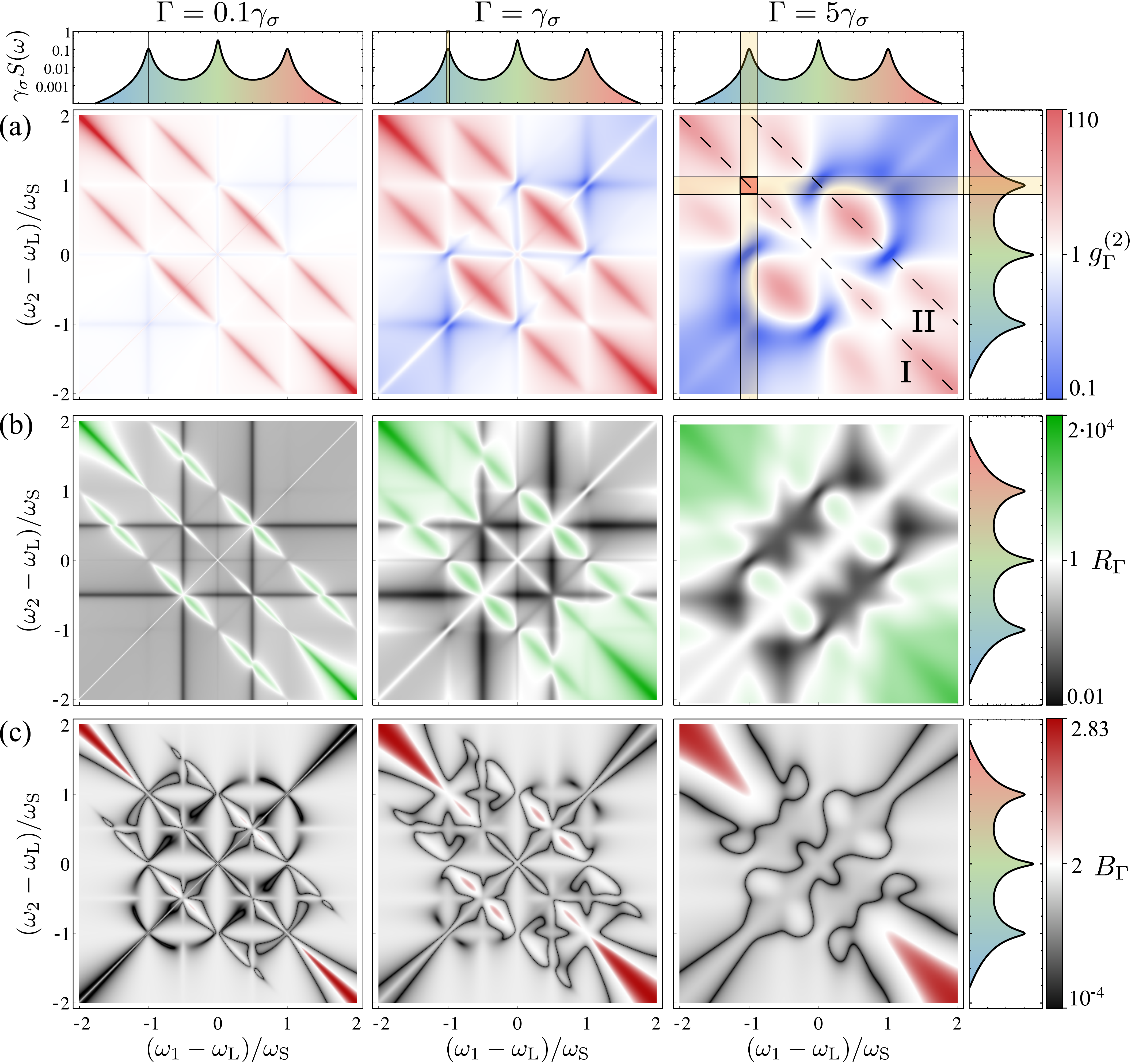}
  \caption{Landscapes of correlations in the frequency domain for
    three different filter linewidths. (a)
    $g^{(2)}_\Gamma(\omega_1,\omega_2)$, (b)
    $R_\Gamma(\omega_1,\omega_2)$ and (c)
    $B_\Gamma(\omega_1,\omega_2)$. In (b) [resp.~(c)], the color code
    is such that green [resp.~red] violates the CSI [resp.~BI] and
    thus corresponds to genuine quantum correlations between the
    detected photons in the corresponding energy windows, while black
    and white do not (with white maximizing the inequality). The
    violation originates from the emission that involves virtual
    states. Dashed lines I and II in (a) are the cuts in the frequency
    domain along which curves in Fig.~\ref{fig:3A} are calculated. The
    spectra on the axes show which frequency windows are
    correlated. Parameters are the same as in Fig.~\ref{fig:1}. An
    animation of these landscapes as a function of the detector
    linewidth is provided in the Supplemental Material. }
\label{fig:2}
\end{figure*}

\section{Results}

At this point, we have set the stage to fully characterize the
quantumness of the emission in terms of violation of the CSI and BI
spanning over all the frequencies of emission and windows of
detection. Note the considerable improvement as compared to the
mode-correlation approach, since a continuum of frequencies in windows
of arbitrary sizes can now be investigated without assumptions on the
order of emission.  Figure~\ref{fig:2} shows three correlation
landscapes in the frequency domain depicting the value of
$g^{(2)}_\Gamma(\omega_1,\omega_2)$, $R_\Gamma(\omega_1,\omega_2)$ and
$B_\Gamma(\omega_1,\omega_2)$ for three different values of the
detector linewidth in an otherwise identical configuration. An
animation of the full landscapes of correlations as a function of the
linewidth of filtering is provided in the Supplemental Material. It
immediately comes across that the quantum character of the emission,
where the inequalities are violated, is structured along three
antidiagonals. In particular, the anticorrelation
$g^{(2)}_\Gamma(\omega_1,\omega_2)<1$ (corresponding to blue areas in
Fig.~\ref{fig:2}(a)), is a CSI violation in time
when~$\omega_1=\omega_2$ and therefore corresponds to a non-classical
effect~\cite{loudon_book00a}. It makes no such guarantee, however, of
a genuine quantum nature when~$\omega_1\neq\omega_2$, and could in
fact even be produced by a classical
emitter~\cite{arXiv_gonzaleztudela14a}. The corresponding CSI
violation in time in this case is
$\left[g^{(2)}_\Gamma(\tau,\omega_1,\omega_2)\right]^2>
g^{(2)}_\Gamma(0,\omega_1,\omega_1)g^{(2)}_\Gamma(0,\omega_2,\omega_2)$, which we study at zero time delay $\tau=0$. This shows the necessity to
turn to such tests to assess the quantum character of an emitter by
frequency filtering. The regions where they are violated indeed
correspond not to frequency antibunching but, on the opposite, to
frequency bunching in the two-photon correlation spectrum. The
reason for this lies in the nature of the violation, with
cross-correlations being higher with respect to auto-correlations than
permitted by classical physics. Physically, the anti-diagonals where this happens
are precisely those where two-photon emission occurs in a ``leapfrog
process''~\cite{gonzaleztudela13a}, i.e., a jump over the intermediate
real state by involving a virtual state instead. This generates the state~$\ket{11}$ that, fed to beam splitters,
generates the maximally entangled state that optimizes the
violation. The antidiagonal, line I, corresponds to transitions from
$\ket{+}$ to $\ket{+}$ or from $\ket{-}$ to $\ket{-}$, two rungs
below, as is sketched in Fig.~\ref{fig:1}(b), thus satisfying
$\omega_1+\omega_2=0$.  Line II and its symmetric correspond to
transitions from $\ket{+}$ to $\ket{-}$ and from $\ket{-}$ to
$\ket{+}$, respectively, satisfying $\omega_1+\omega_2 = \pm
\omega_\mathrm{S}$.  The CSI and BI are less, or are not, violated
whenever the intermediate rung intersects a real state, as seen by the
fact that the green (for $R_\Gamma$) and red (for $B_\Gamma$) regions are depleted
or pierced when intersecting the sidebands $\pm
\omega_\mathrm{S}$. This is particularly important since previous
studies focused precisely on correlations between real transitions,
i.e., between peaks, such as indicated by the red square in the
rightmost panel of Fig.~\ref{fig:2}(a). Instead, the exact treatment
shows that these are detrimental to the effect, that is optimum when
involving virtual states, since these are the vector of quantum
correlations. It is easy to prove, from the closed form expression
Eqs.~(6--8) in Ref.~\cite{gonzaleztudela13a}, that a single-mode
emitter with no dressing (here by the laser) never violates the CSI,
regardless of frequencies and detection widths. The same was checked
numerically for the case of BI. Notably, this is true even if the
emitter is a two-level system and exhibits perfect antibunching,
$g^{(2)}(\tau=0)=0$. All this evidence confirms that CSI and BI
violations are rooted in the quantum dynamics that involves a virtual
state in a collective de-excitation in the quantum ladder of the
dressed states.

\begin{figure}[]
  \centering
  \includegraphics[width=.85\linewidth]{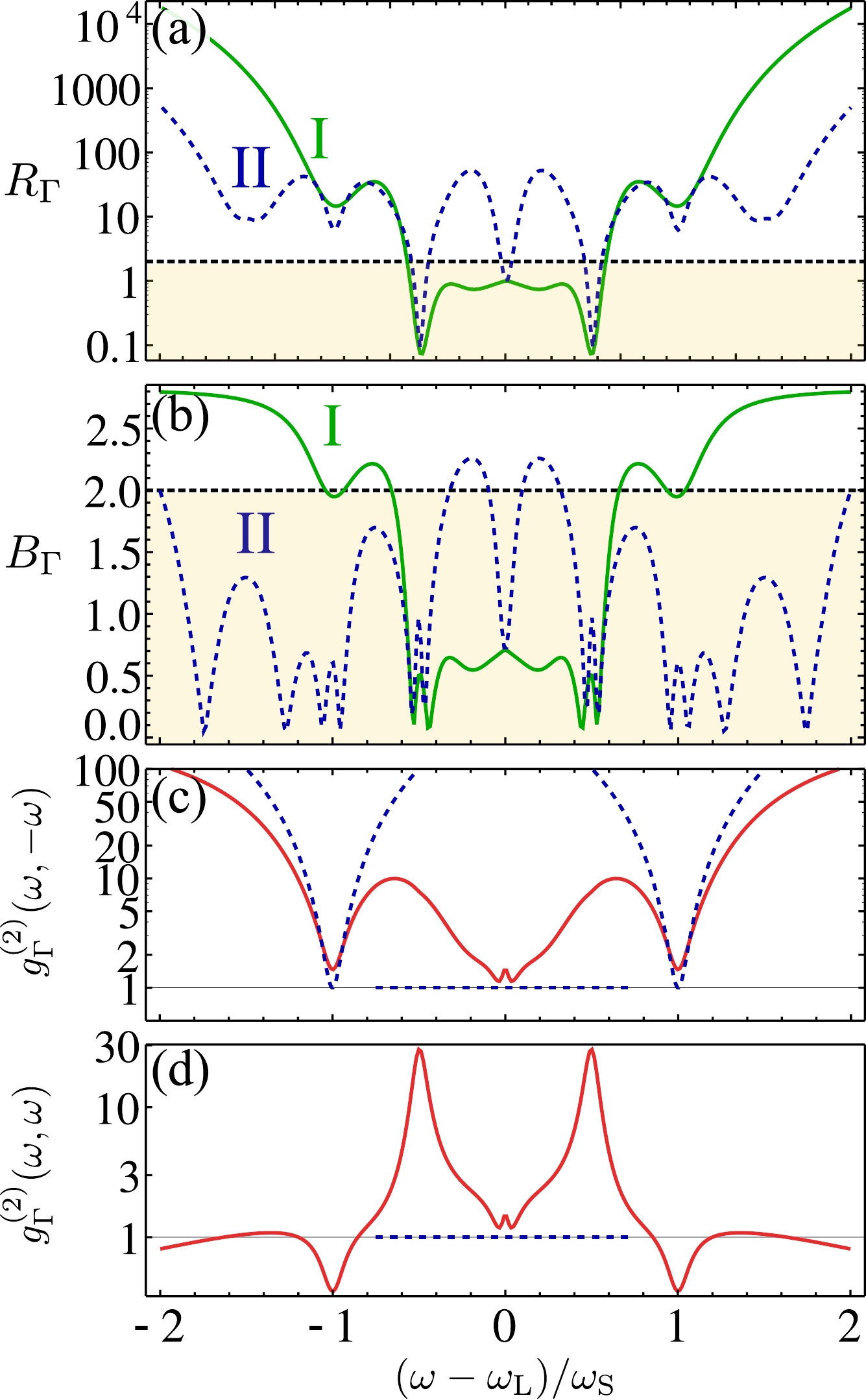}
  \caption{(Color online) (a--b): Cuts of $R_\Gamma$ (a) and
    $B_\Gamma$ (b) along the lines~I $(\omega,-\omega)$ and~II
    $(\omega,\omega_\mathrm{S}-\omega)$ of
    Fig.~\ref{fig:2}(a). (c--d): photon-correlation
    $g^{(2)}_\Gamma(\omega,\pm\omega)$ computed exactly (solid red) or
    through the usual multiple-mode approximation (dashed blue). In
    panel (d), the absence of the latter curve in some domains
    correspond to values which are, incorrectly, exactly zero (the
    vertical axis is in log-scale).}
  \label{fig:3A}
\end{figure}

A more quantitative reading of these results is given in
Fig.~\ref{fig:3A}(a--b), that shows slices in the landscapes along
lines I and II of the rightmost panel of Fig.\ref{fig:2}(a).  The
quantum correlations violating the CSI are found in the side peaks and
beyond, being larger the farther from the peaks. The same feature is
present in the BI violation, which furthermore tends to the maximum
value allowed~$B_\Gamma = 2\sqrt{2}$.
\begin{figure}[]
  \centering
  \includegraphics[width=\linewidth]{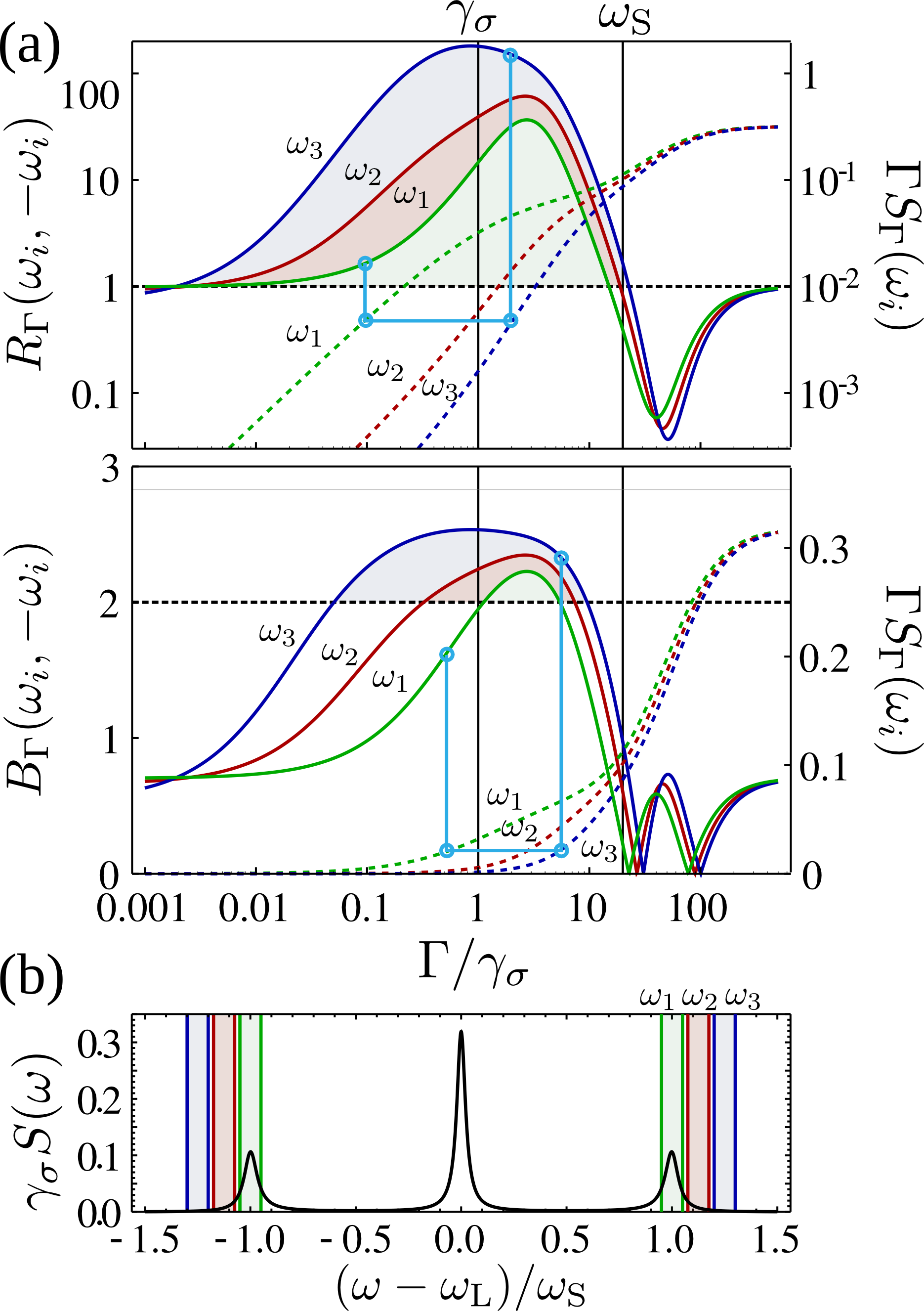}
  \caption{(Color online) (a) Solid lines:
    $R_\Gamma(\omega_i,-\omega_i)$ (top panel) and
    $B_\Gamma(\omega_i,-\omega_i)$ (bottom panel) as a function of the
    detector linewidth for the three set of frequencies
    $(\omega_i,-\omega_i)$, $i\in {1,2,3}$ depicted in panel
    (b). Dashed lines: Amount of signal $\Gamma S_\Gamma(\omega)$ that
    can be collected for the corresponding filter linewidth. Blue
    points illustrate how two configurations with the same amount of
    collected signal can yield different degrees of violation. (b)
    Resonance fluorescence spectrum, this time in linear scale,
    displaying the characteristic Mollow triplet and three sensors
    with linewidth $\Gamma=2\gamma_\sigma$ centred at the frequencies
    used for panel (a): $\omega_1=\omega_\mathrm{S}$,
    $\omega_2=1.125\,\omega_\mathrm{S}$ and $\omega_3=1.25 \,
    \omega_\mathrm{S}$. Parameters are the same as in
    Fig.~\ref{fig:1}.}
  \label{fig:3B}
\end{figure}
Figure~\ref{fig:3A}(c--d) show~$g_\Gamma^{(2)}(\omega,\omega)$
and~$g^{(2)}_\Gamma(\omega,-\omega)$---that can be used to derive
$R_\Gamma(\omega,-\omega)$ and an approximation of
$B_\Gamma(\omega,-\omega)$~\footnote{An approximation for
  $B_\Gamma(\omega,-\omega)$ can be obtained by dropping the last two
  terms of the numerator in
  Eq.\eqref{eq:ansari}~\cite{ansari97a}. This approximation can be
  applied as long as there are no second-order transfer of photons
  between the modes, but these terms could be important in other
  cases. We have verified that they are indeed negligible in our
  configuration.}--- as calculated exactly (solid red
lines)~\cite{delvalle12a,gonzaleztudela13a} and through the auxiliary
multi-mode approximation used in previous works (dashed
blue)~\cite{schrama92a,nienhuis93a}.
In the multi-mode approximation, the estimation is local around the
peaks, that is, at $\omega/\omega_\mathrm{S}=\pm1$ and~$0$ (dotted
vertical lines), where it is seen to be fairly accurate indeed,
although not numerically exact. It can still lead to qualitative
error, e.g., the autocorrelation at the sidebands is exactly zero in
this approximation, predicting arbitrary violation of the CSI even
when it is obeyed and an unphysical violation of the BI. A violation
of the Bell's inequality in these terms was predicted in
Ref.~\cite{joshi91b}. However, the violation was considered
ill-defined due to the perfect antibunching of the
sidebands. Furthermore, these expressions are found in limiting cases
for the filter linewidths: either $\Gamma \ll \gamma_\sigma \ll
\Omega$ or $\gamma_\sigma \ll \Gamma \ll \Omega$. Both assume that the
peaks are well separated to allow for the multiple-mode approximation.
They predict no CSI or BI violation for narrow filters, which is
ultimately verified although it is for values of the detector
linewidth so small that they are unphysical. Solid lines in
Fig.~\ref{fig:3B}(a) show the dependence of $R_\Gamma$ and $B_\Gamma$
on the detector linewidth $\Gamma$ for the three sets of frequencies
$(\omega_i,-\omega_i)$, $i\in {1,2,3}$ depicted in
Fig.~\ref{fig:3B}(b).  For the already extremely small value of
frequency windows $\Gamma=0.1\gamma_\sigma$, the CSI and BI can be
violated, in contradiction with the prediction of the multiple-mode
approximation.

There are mainly three regimes of frequency correlations: narrow
filters, peak filtering and overlapping windows.  While narrow filters
better define the structure, as can be seen in Fig.~\ref{fig:2}, they
also correspond to longer times of integration due to the
time-frequency uncertainty and thus average out the correlations. A
maximum is found when filtering in windows of the order of the peak
linewidth or above, which is a welcomed result for an experimentalist.
The overlap of the filters marks a change of trend in all the curves,
due to a competition between various phenomena involving, for
instance, various transitions as well as averaging over different
types of interferences.  Dashed lines in Fig.~\ref{fig:3B}(a) show the
value of $\Gamma S_\Gamma(\omega)$ corresponding to the amount of
signal that can be collected with a detector of linewidth $\Gamma$ at
the frequency $\omega$~\cite{delvalle12a}. This way, one can easily
compare, for a given amount of available signal, the different degrees
of violation which are accessible simply by selecting the frequency
and the window of the detector appropriately.  Since such correlations
are useful for technological purposes, the ability to compute the
entire landscape of frequency correlations becomes helpful for
optimizing quantum information processing.  Correlations along line I
of the map arise from a well defined family of virtual processes, from
which sideband correlations have been shown to be just a particular,
and in fact also a detrimental case.  By positioning the filters away
from the sidebands and increasing the frequency window of detection,
it is possible to distill light showing stronger quantum correlations
without paying any price on the signal.  \vspace{10pt}

\section{Conclusions}
We have shown how to evidence and optimize CSI and BI
violations between photons resolved in frequency from a quantum
source, with no constrains nor approximations from the theoretical
description. Maximum violation is to be found not when correlating
peaks in the spectrum, as previously thought, and thus linked to
transitions between real states, but when involving virtual processes
in the quantum dynamics. These results show the potential of frequency
correlations to engineer quantum correlations, and could be applied
towards the design of optimum quantum information processing devices.
\section{Acknowledgements}
We acknowledge the IEF project SQUIRREL (623708), the Spanish MINECO
(MAT2011-22997, FPI \& RyC programs) and the
ERC PolaFlow.
\bibliography{Sci,books,arXiv}

\end{document}